\documentclass[prl,twocolumn,amsmath,amssymb,aps,]{revtex4-1}
\usepackage[utf8]{inputenc} 
\usepackage{graphicx}
\usepackage{adjustbox}
\usepackage{dcolumn}
\usepackage{bm}
\usepackage{caption}
\usepackage{subcaption}
\usepackage{comment}
\usepackage{hyperref}
\usepackage{float}
\hypersetup{
     colorlinks   = true,
     linkcolor    = blue,
     citecolor    = blue
}
\usepackage[all]{hypcap}
\begin{document}
\preprint{APS/123-QED}

\title{$\mathbb{Z}_3$ parafermionic zero modes without Andreev backscattering from the $2/3$ fractional quantum Hall state}

\author{Yahya Alavirad$^1$}
\author{David Clarke$^1$}
\author{Amit Nag$^1$}
\author{Jay D. Sau$^1$}

\affiliation{
$^1$Department of Physics, Condensed Matter theory center and the Joint Quantum Institute, University of Maryland, College Park, MD 20742.
}

\date{\today}

\begin{abstract}
Parafermionic zero modes are a novel set of excitations displaying non-Abelian statistics somewhat richer than that of Majorana modes. These modes are predicted to occur when nearby fractional quantum Hall edge states are gapped by an interposed superconductor. Despite substantial experimental progress, we argue that the necessary crossed Andreev reflection in this arrangement is a challenging milestone to reach. We propose a superconducting quantum dot array structure on a fractional quantum Hall edge that can lead to parafermionic zero modes from coherent superconducting forward scattering on a quantum Hall edge. Such coherent forward scattering has already been demonstrated in recent experiments. We show that for a spin-singlet superconductor interacting with loops of spin unpolarized $2/3$ fractional quantum edge, even an array size of order ten should allow one to systematically tune into a parafermionic degeneracy.
\end{abstract}

\maketitle

\emph{Introduction.}---Theoretical understanding and experimental realization of non-Abelian anyons has attracted considerable attention in the past few years. This surge of interest can be largely attributed to potential application of such systems as building blocks for topological quantum computers~\cite{Nayak08}. Majorana zero modes (MZMs)~\cite{Kitaev01,Alicea12,Leijnse12,Beenakker13,Stanescu13,Franz15} provide the simplest and experimentally the most promising example of non-Abelian anyons. So far, most of the effort in searching for non-Abelian anyons has been focused on MZMs. Following a series of theoretical proposals~\cite{Sau10,Alicea10,Lutchyn10,Oreg10}, suggestive experimental evidence of MZMs has been observed in semiconductor/superconductor heterostructures~\cite{Exp1,Exp2,Exp3,Exp4,Exp5,Exp6,Exp7}. Despite their fascinating properties MZMs are non-Abelian anyons of the Ising ($\mathbb{Z}_2$) type. Universal quantum computation cannot be implemented using braiding of the $\mathbb{Z}_2$ anyons alone. Therefore, searching for a computationally richer set of anyons seems necessary.

 Parafermionic zero modes (PZMs)~\cite{Alicea16,Fendley12,Fradkin80} (also known as fractional Majorana modes) provide an example of such computationally rich (still not universal) anyons. They can be thought of as $\mathbb{Z}_n$ generalizations of MZMs. Similar to MZMs, $\mathbb{Z}_n$ PZMs are associated with $n$ fold degeneracy of the ground state that is robust to all local perturbations. Due to fundamental restrictions set on possible topological phases in strictly one dimensional systems\cite{Fidkowski11,Turner11}, PZMs cannot exist in isolation. However, recently it was realized that boundaries of two dimensional systems can circumvent these restrictions. It was explicitly shown that PZMs emerge at the one dimensional boundary of two counter-propagating fractional quantum Hall (FQH) edges coupled with superconducting contacts~\cite{Clarke13,Linder12,Cheng12}. These setups greatly resemble the canonical proposal used to realize MZMs\cite{Fu08}, with the role of topological insulator played by a pair of opposite-chirality FQH states. All of the existing proposals (involving superconductors) require two main ingredients, induced superconductivity via coupling FQH edge state to a superconductor and crossed Andreev tunneling between two edges. The first requirement has already been achieved in experiments~\cite{Amet16,Wan15}. However, the second requirement is likely to be difficult to achieve experimentally due to the disruption of FQH edge states placed adjacent to a superconductor. This is because strong coupling to the superconductor is likely to change the density in the surrounding 2D electron gas, pushing the FQH edges away from the superconductor. The amplitude of quasiparticle tunneling between the edges would then be reduced by the increased distance and the Fermi wave-vector in the intervening superconducting region.

 \begin{figure}[H]
\centering
\includegraphics[width=.9\columnwidth,keepaspectratio]{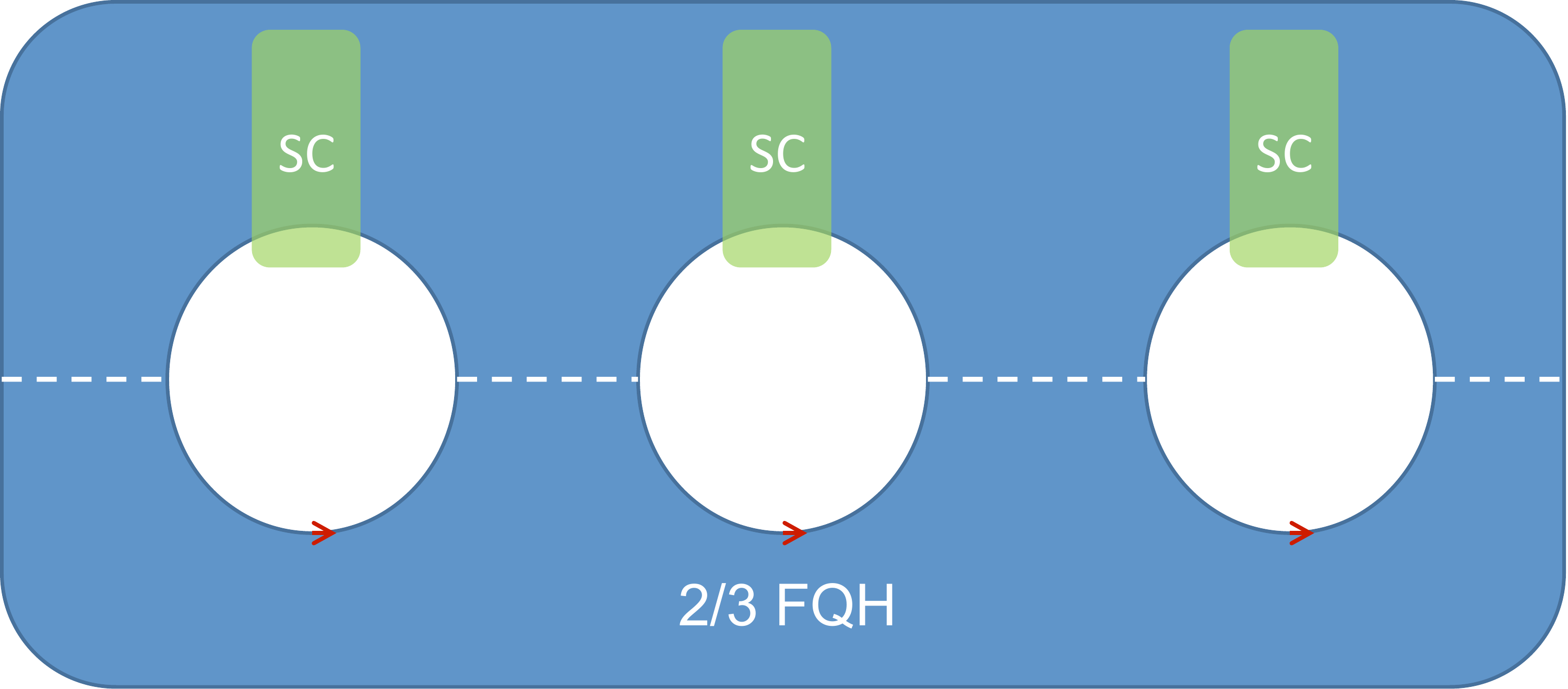} 
\caption{Top view of the system, comprised of a linear array of superconductors coupled to loops of FQH edge states, different loops are connected via quasiparticle hopping \label{fig:pic1.pdf}}
\end{figure}

In this letter we propose a practical scheme to realize $\mathbb{Z}_3$ PZMs from the spin unpolarized $2/3$ fractional quantum Hall state using superconducting contacts without cross-Andreev tunneling, which can be realized in present experiments. Our system is comprised of a linear array of FQH edge loops; each one of these loops is coupled to a superconductor through proximity effect, and separate loops are connected via quasiparticle tunneling.  The strength of the quasiparticle tunneling can be controlled with a gate voltage. A top view of this setup is shown in Fig.~\ref{fig:pic1.pdf}. We use a combination of analytical and numerical methods to study this model and show that for realistic values of parameters, at relatively small chain lengths (order ten loops) this system can be tuned to a topological phase hosting $\mathbb{Z}_3$ PZMs.

\emph{Model.}---We begin by studying a single loop coupled to a superconductor. Assuming $SU(2)$ symmetry, the effective Hamiltonian describing the charge part of the $\nu=2/3$ FQH edge state is given by the following chiral boson theory\cite{Wen04,Wen95},
\begin{align}\label{1}
H_{edge}=\int_0^L dx [\frac{u}{4 \pi \nu}(\partial_x \varphi(x))^2 -\frac{u m_\mu}{2L} \partial_x \varphi(x)]
\end{align}
where $L$ is the length of the loop, $u$ is the mode velocity, $ m_\mu$ is the gate controlled dimensionless chemical potential (as opposed to actual chemical potential $\mu=\frac{u m_\mu}{2L}$) and $\varphi(x)$ is the chiral boson field that is defined in terms of the charge density operator as $\rho=\frac{1}{2\pi}\partial_x \varphi(x)$. The $\varphi(x)$ field obeys the commutation relation $ [\varphi(x),\partial_y\varphi(y)]=2 i \pi \nu \delta(x-y)$.
Using this relation we can write the charge $2/3$ spinless quasiparticle creation operator as $e^{i \varphi(x)}$ and charge $2$ Cooper pair creation operator as $e^{3i \varphi(x)}$. 
The neutral mode, which does not couple to the SC and is expected to be non-degenerate and gapped will be ignored in the rest 
of this work. The edge Hamiltonian $H_{edge}$ can be diagonalized \cite{Wen90} by mode expanding $\varphi(x)$ as
\begin{align}\label{me}
&\varphi(x)=\frac{2\pi \hat{n}\nu x}{L}+\hat{\varphi}_0+\nonumber\\&\sum_{k= 0}^{K_{max}}\left[-i\sqrt{\frac{\nu}{k}}a^\dagger_ke^{2\pi i k x/L}+i\sqrt{\frac{\nu}{k}}a_ke^{-2\pi i k x/L}\right]
\end{align}
where $a_k , a^\dagger_k$ are the $k$th momentum boson creation and annihilation operator for $k \in \mathbb{N}$, $\hat{\varphi}_0 , \hat{n}$ are the zero mode phase and  number operators, respectively, and $K_{max}$ is the momentum cutoff. 

Now we can rewrite Eq.\eqref{1} as,
\begin{align}\label{h2}
H_{edge}=\frac{u\pi\nu}{L} \big( \hat{n}^2- m_\mu \hat{n} + \frac{2}{\nu} \hat{P}  \big) +const
\end{align}
where $\hat{P}=\sum_{k=0}^{K_{max}}  k \hat{a}^{\dagger}_{k}\hat{a}_{k}$ is the total momentum operator. When the dimensionless chemical potential $ m_\mu$ is tuned by a gate voltage to integer values the spectrum is invariant under changing $n=m$ to $n=-m+ m_\mu$. For odd $ m_\mu$ this translates into a two fold ground state degeneracy. This degeneracy survives the addition of superconductivity and will play a crucial role later on.

The effect of the superconductor on a single loop can be modeled by the Hamiltonian
\begin{align}
H_{sc}=\int_0^L dx \frac{\Delta(x)}{L} e^{-\frac{2\pi i x  m_\mu}{L}}e^{3i \varphi(x)} +h.c
\end{align}
describing the tunneling of Cooper pairs to and from the superconductor.
Here $\Delta(x)$ corresponds to the position-dependent Cooper pair tunneling amplitude and $e^{-\frac{2\pi i x  m_\mu}{L}}$ is the phase factor taking into account the chemical potential mismatch between the FQH edge and the (grounded) superconductor. Fourier transforming $\Delta(x)=\sum_{k} \Delta(k) e^{\frac{2\pi i kx}{L}}$ and mode expanding $\varphi(x)$ (as in Eq.~\eqref{me}) allows us to write the only nonzero matrix elements of $H_{sc}$:
\begin{align}\label{me1}
&\langle n_0\pm3 ,\{m_k\}|H_{sc}|n_0 ,\{n_k\}\rangle=\sum_{k} \Delta(k) \delta_{(\Delta E\pm 3k)}\\ \nonumber
&\times \langle n_0\pm3 ,\{m_k\}| e^{\pm3i \varphi(0)} |n_0 ,\{n_k\}\rangle,
\end{align}
where $\Delta E = E(n_0\pm3,\{m_k\})-E(n_0,\{n_k\})$ is the energy difference between the initial and the final state, and $E=\frac{u\pi\nu}{L}(n^2- m_\mu n+3 P)$ is the bare edge energy of each state in accordance with Eq.~\eqref{h2}. Equation \eqref{me1} implies that the special case of uniform superconductivity $\Delta(x)=\Delta_0$ leads to the additional conserved quantity $H_{edge}$, as $\left[H_{sc}, H_{edge}\right]=0$. Though convenient, this symmetry is not generic and is therefore not used in this letter.

\begin{figure}[tb]
\centering
\includegraphics[width=\columnwidth,keepaspectratio]{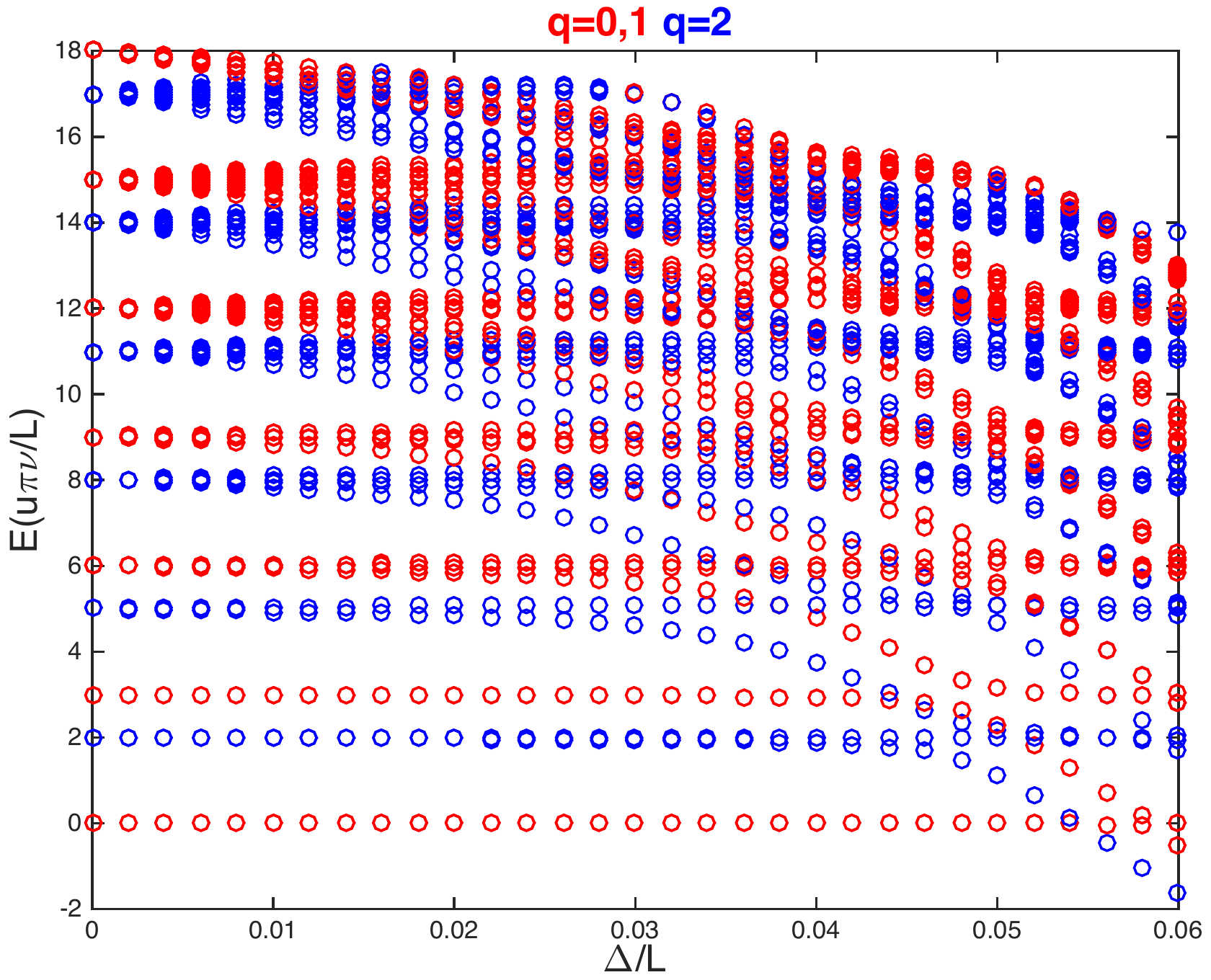}
\caption{Low-lying spectrum for ``pseudo-point-like'' superconductivity as a function of $\frac{\Delta}{L}$. Here $ m_\mu=1$, $K_{max}=4$ and $q$ is the fractional charge modulo three $q=\mathrm{mod}(n,3)$. All red circles are two fold degenerate, blue circles are non-degenerate.\label{fig:pic2.pdf}}
\end{figure}

The inclusion of $H_{sc}$ reduces the conservation of fractional charge $n$ to conservation of $q=\mathrm{mod}(n,3)$, which only takes three values $q=0,1,2$. Using this restriction we can divide the system into three independent charge sectors with $q=0,1,2$. For integer $ m_\mu$, the $n=3m$ to $n=-3m+ m_\mu$ symmetry of the non-superconducting edge now translates into degeneracy of two of the charge sectors, for example at $ m_\mu=1$ the two sectors $q=0,1$ will become degenerate.

Using Eq.\eqref{me1} we can numerically calculate the spectrum of a single loop $H=H_{edge}+H_{sc}$. We set $ m_\mu=1$ and assume a ``pseudo-point-like" superconductivity such that $\Delta(k)=\Delta$ for $ |k| \leqslant K_{max}$, where $K_{max}$ is the momentum cutoff defined in Eq.\eqref{me}. The low-lying part of the spectrum is plotted in Fig.~\ref{fig:pic2.pdf}. This plot shows that the ground state is separated from the rest of the spectrum by a gap for a finite range of $\Delta$. However, the ground state degeneracy (between $q=0,1$) remains two-fold with $q=2$ split.

\emph{Effective Hamiltonian.}---In the absence of superconductivity and for odd values of the dimensionless chemical potential ($ m_\mu=2n-1$), the  ground-state energy of the system is twofold degenerate and is separated from the excited states by a gap for a range of $\Delta$.  The two ground states can be labeled by fractional charge $q=0,1$. Therefore as long as we choose $\Delta$ in this range and restrict the ratio of hopping amplitude to the energy gap $t/\Delta E$ to be small, we can apply a Schrieffer-Wolff transformation to obtain an effective Hamiltonian defined in the Hilbert space spanned by ground states of single loops. This emergent Hilbert space has only two states per site ($q=0,1$) and therefore can be thought of as a chain of spin $1/2$ particles, where the states with spin up/down correspond to the single loop ground states with $q=1,0$.

To calculate the low-energy effective Hamiltonian, we start with the Hamiltonian describing quasiparticle hopping between different loops,
\begin{align}\label{t1}
H_{tunnel}=t\sum_i e^{i \varphi_i(L/2)} e^{-i\varphi_{i+1}(0)} + H.c.
\end{align}
Note that $e^{i \varphi_i(x)}$ is an anyonic operator and has nontrivial commutation relation with other anyonic operators. For different sites we can write,
\begin{align}
e^{i \varphi_i(x)}e^{i \varphi_j(x')}=e^{i \varphi_j(x')}e^{i \varphi_i(x)} e^{i \pi \nu \text{sgn}(j-i)}.
\end{align}
Using a generalized Jordan-Wigner string we can define the new field variables $\tilde{\varphi}(x)$,
\begin{align}
e^{i \varphi_i(x)}=e^{i \pi\nu \sum_{j<i} \tilde{n}_{j}}e^{i \tilde{\varphi}_i(x)}.
\end{align}
The advantage of these new field variables is that they commute trivially between different sites $[e^{i \tilde{\varphi}_i(x)},e^{i \tilde{\varphi}_j(x')}]\propto \delta_{i,j}$, and therefore act strictly on the local loop Hilbert space. Now we can rewrite $H_{tunnel}$ as,
\begin{align}
H_{tunnel}=t\sum_i e^{i \tilde{\varphi}_i(L/2)} e^{-i\pi\nu \tilde{n}_{0,i}}e^{-i\tilde{\varphi}_{i+1}(0)} + H.c. .
\end{align}
To second order in perturbation theory we can write the low energy effective Hamiltonian as,
\begin{align}\label{pert}
H_{eff}=&PH_{tunnel}P - PH_{tunnel}\frac{(1-P)}{H_0}H_{tunnel}P \\ \newline \nonumber
&+O(t^3/\Delta E^2)
\end{align}
where $P$ is the single loop ground state projection operator defined as $P=\sum_i (|q=i\rangle\langle q=i|)$, and $H_0$ is the shifted single loop Hamiltonian (shifting to set ground state energy to zero).

\begin{figure}
\vspace*{0.1in}
\centering
\includegraphics[width=\columnwidth,keepaspectratio]{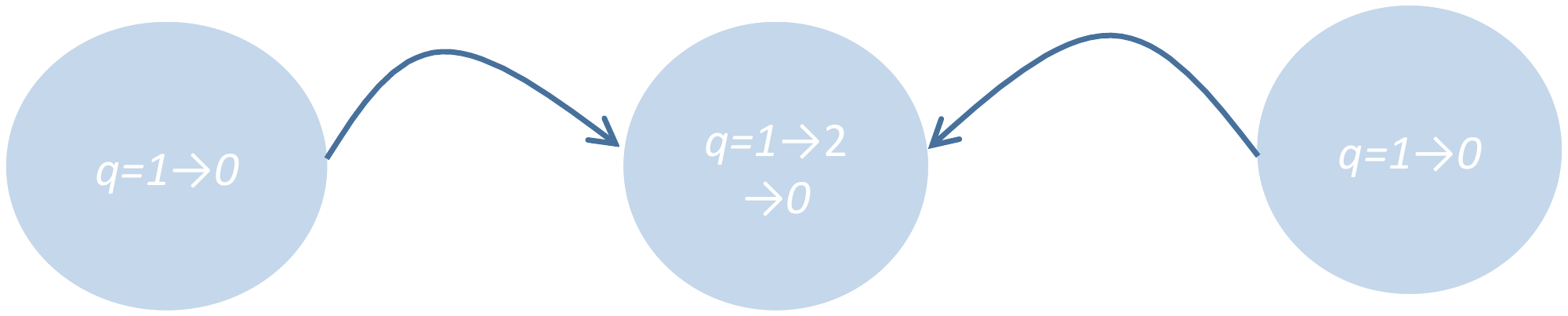} \caption{$\sigma^-_{i-1} \sigma^-_{i} \sigma^-_{i+1}$ term as a second order process in perturbation theory. $q$ is the fractional charge modulo three. Two fractional charge are tunneled to the middle site, one from each neighbour \label{fig:pic3.pdf}}
\end{figure}

Putting everything together we get (Details of this calculation can be found in the supplementary material, here we just quote the final results),
\begin{align}\label{sm}
H_{eff}=&\sum_i \Big[ (t \alpha_0 e^{i\pi/3}- t^2 \alpha_1  e^{-2i\pi/3})  \sigma^+_i\sigma^-_{i+1} \\ \newline \nonumber
& -t^2 e^{2i\pi/3}\gamma \sigma^+_{i-1}\sigma^-_{i+1} -t^2\beta \sigma^z_i\sigma^z_{i+1} \\ \newline \nonumber
&+t^2 \lambda \sigma^+_{i-1} \sigma^+_{i} \sigma^+_{i+1} \Big] + H.c. +O(t^3/\Delta E^2)
\end{align}
where $\sigma$'s are the usual Pauli matrices. Note that all terms in the Hamiltonian conserve fractional charge (spin) modulo three and are also $\mathbb{Z}_2$ symmetric under $\sigma^z\rightarrow -\sigma^z$ , this $\mathbb{Z}_2$ symmetry can be associated with the $\varphi\rightarrow -\varphi+\frac{2\pi m_\mu\nu x}{L}$ symmetry of the original Hamiltonian (that we have already discussed). Note that presence of the term $t^2 \lambda \sigma^+_{i-1} \sigma^+_{i} \sigma^+_{i+1} $ requires nonzero superconductivity, since without $\Delta$ fractional charge (spin) has to be conserved. As seen in Fig.~\ref{fig:pic3.pdf}, this term, which arises at second order in tunneling, requires $H_{sc}$ so that  q=3 may be converted to q=0 by the removal of a Cooper pair.

\begin{table}
\centering
\begin{tabular}{|l|l|l|l|l|}
\hline
\begin{tabular}[c]{@{}l@{}}Chain \\ length\end{tabular} & \begin{tabular}[c]{@{}l@{}}Ground \\ state energy\end{tabular} & \begin{tabular}[c]{@{}l@{}}1st excited \\ state energy\end{tabular} & \begin{tabular}[c]{@{}l@{}}2nd excited \\ state energy\end{tabular} & \begin{tabular}[c]{@{}l@{}}3rd excited \\ state energy\end{tabular} \\ \hline
10           & -6.818                                                        & -6.696                                                              & -6.696                                                              & -6.149                                                              \\ \hline
40           & -29.681                                                               & -29.677                                                                    & -29.677                                                                    & -29.384                                                                    \\ \hline
100          & -75.524                                                       & -75.524                                                            & -75.524                                                            & -75.267                                                              \\ \hline
\end{tabular}
\caption{DMRG calculation results for ``pseudo point-like" superconductivity(defined earlier) at $t=1$, $\Delta/L=0.046$ and momentum cutoff $K_{max}=4$}\label{tb2}
\end{table}

\emph{Analysis.}---In the absence of the superconductivity ($\sigma_z$ non-conserving terms), the conservation of $\sigma_z$ ensures a gapless state with low-energy Luttinger liquid Hamiltonian where $\sigma_z\sim \nabla \phi$. In this Luttinger liquid description, the superconducting term  $t^2 \lambda \sigma^+_{i-1} \sigma^+_{i} \sigma^+_{i+1}$ is represented as $g \cos (3\theta)$, which converts the Luttinger liquid to a Sine-Gordon model. For the correct choice of parameters the superconductivity induced term $g \cos (3\theta)$ becomes perturbatively relevant \cite{giamarchi} and gaps out the system into a topological phase with a $\mathbb{Z}_3$ parafermionic degeneracy, where each ground state corresponds to the phase $\theta$ being locked at one of the three minima of the $\cos(3\theta)$ term. To check whether this degeneracy occurs in a our system (i.e. the Hamiltonian in Eq. \eqref{sm}) with realistic values of the parameters, we numerically study the Hamiltonian in Eq.\eqref{sm} using the DMRG method. DMRG calculations were performed using the ITensor C++ library\cite{ITensor}. Sample results of this calculation are shown in Table.~\ref{tb2}. These results confirm existence of a three-fold degeneracy for reasonable parameters. The degeneracy is weakly split for small chain lengths $N\approx 10$ and is more pronounced at longer lengths, as expected from a true topological degeneracy. We further emphasize that despite our calculation being done for a relatively small momentum cutoff $K_{max}=4$ (chosen for calculational simplicity), we expect our results to hold quite generically, independent of the value of $K_{max}$.

An alternative interesting limit is that of ``true" point-like superconducting contacts (as opposed to pseudo point like defined before), that is $\Delta(k)=\Delta~\forall k$. This limit is particularly appealing, as in this case analytical results may be obtained for large values of $\Delta$. Following the formalism developed in Ref.\cite{Ganeshan16}, we show (with details in the Supplementary Material) that in the large $\Delta$ limit the system is described by set of decoupled harmonic oscillators, and that in this limit all three fractional charge sectors become degenerate. Analogous to the previous case (small $\Delta$) as long as $t/\Delta E$ is small, we can use Wolff transformation to find an effective Hamiltonian. In this case the effective Hilbert space of each site is three dimensional (corresponding to three fractional charge sectors) and can be though of as a three state clock model. In this limit, it's useful to define,

\begin{align}
\alpha_{2j-1} = \frac{e^{-i\varphi_{j}(0)}}{A(0)} ; \alpha_{2j} = \frac{e^{-i\varphi_{j}(L/2)}}{A(L/2)}
\end{align}
where $A(x)=(\prod_i\langle q=i|e^{-i\varphi(x)}|q=i+1\rangle)^{1/3}$ is a normalization factor. Within the effective Hilbert space these operators are the usual parafermionic operators, that is $\alpha^3_j=1$ and $\alpha_j \alpha_{j'}=\alpha_{j'}\alpha_j e^{i \frac{2\pi}{3}sgn(j'-j)}$. Using these variables and the Hamiltonians in Eqs.\eqref{t1} and \eqref{pert} we arrive at,
\begin{align}\label{sm2}
H_{eff}= t \sum_i (A^*(L/2)A(0) \alpha^{\dagger}_{2j} \alpha_{2j+1}+ H.c.) +O(t^2/\Delta E)
\end{align}
in this form presence of parafermionic edge zero modes ($\alpha_1,\alpha_{2N+1}$) is already manifest\cite{Fendley12}. Note that in contrast to Eq.\eqref{sm}, here the calculation has been done to first order in $t$. Using the usual clock model variables \cite{fnt} we can write the Hamiltonian in a more familiar form,

\begin{align}\label{sm3}
H_{eff}=  \sum_i (-J \sigma^{\dagger}_{j} \sigma_{j+1}+ H.c.) +O(t^2/\Delta E)
\end{align}
where $J=t A^*(L/2)A(0) e^{i\pi/3}$ and
\begin{align}
\sigma=\begin{pmatrix}
 1&  & \\
 & \omega & \\
 &  &\omega^2
\end{pmatrix} , \omega=e^{2\pi i/3}
\end{align}
is the ``clock" operator.  In the large but finite $\Delta$ regime, three fold ground state degeneracy of single loops is not exact. We can take this energy difference into account by adding a term $h(\tau_j+\tau_j^{\dagger})$ to Eq.\eqref{sm3} where $h$ is the energy difference of charge sectors $q=0,1$ with the charge sector $q=2$. Estimates for value of $h$ can be found in the supplementary material. Putting everything together we have,
\begin{align}\label{sm4}
H_{eff}=  \sum_i (-J \sigma^{\dagger}_{j} \sigma_{j+1}+ H.c.)+ h(\tau_i+\tau_i^{\dagger}) + O(t^2/\Delta E)
\end{align}
Note that $\sigma_i$ is a non-local operator in the physical system of interest. This is an important point as locality prevents the introduction of a Hamiltonian term proportional to $\sigma_i$. With this constraint and for small values of $h$ ($h$ can be made arbitrarily small by choosing large enough $\Delta$) the Hamiltonian in Eq.\eqref{sm4} is well known to be in a topological phase with three-fold degeneracy\cite{Alicea16}.

\emph{Conclusion.}--- In this work we have considered a linear array of superconducting ``quantum dot"-like holes on a spin singlet $2/3$ fractional quantum Hall sample and showed that for both large and small values of induced superconductivity $\Delta$, this system can be tuned to a topological phase hosting $\mathbb{Z}_3$ PZMs. Unlike earlier proposals used to realize PZMs, our approach does not rely on Andreev back-scattering between two fractional quantum Hall edges. We believe this feature makes our proposal suitable for realization in experiments using ingredients that have already been demonstrated.

This work was supported by the NSF-DMR-1555135, Microsoft and JQI-NSF-PFC.

\bibliographystyle{h-physrev}
\bibliography{library}

\begin{thebibliography}{10}
\bibitem{fnt}
$\alpha_{2j-1}=\sigma_j \prod_{i<j}\tau_i ;
  \alpha_{2j}=-e^{i\pi/3}\tau_j\sigma_j \prod_{i<j}\tau_i $ where
  $\tau=\begin{psmallmatrix}0& 0 & 1\\ 1& 0 & 0\\ 0& 1 & 0 \end{psmallmatrix}$
  is the ``clock shift operator" and $\sigma$ is defined in the main text.
\end{thebibliography}


\begin{centering}
{\hskip .6in \Large \bf Supplementary Material}
\end{centering}

\renewcommand{\thesection}{S\arabic{section}}
\renewcommand{\theequation}{S\arabic{equation}}
\renewcommand{\thefigure}{S\arabic{figure}}
\setcounter{equation}{0}
\setcounter{figure}{0}

\section{Effective spin model parameters}
We start by calculating the first term in Eq.\eqref{pert}, we can graphically represent this term as,
\begin{figure}[H]
\vspace*{0.1in}
\centering
\includegraphics[width=0.6\columnwidth,keepaspectratio]{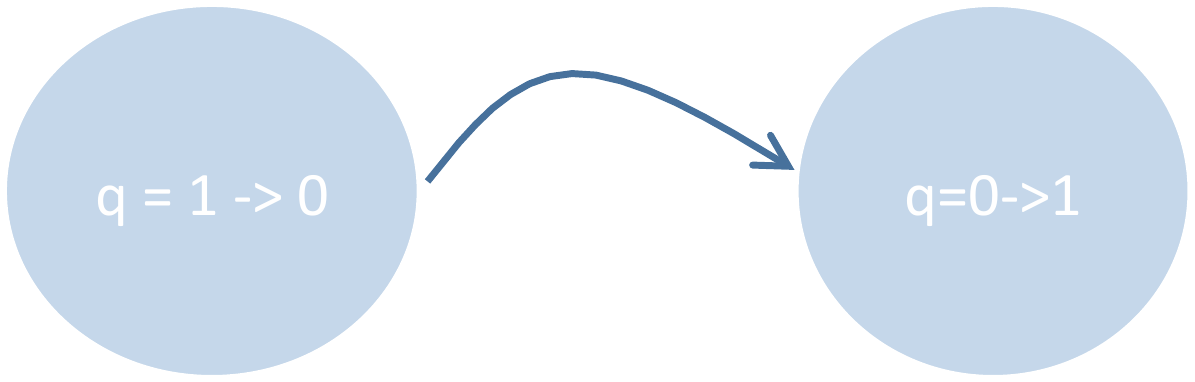} \caption{$\sigma^-_{i} \sigma^+_{i+1}$ term as a first order process in perturbation theory. $q$ is the fractional charge modulo three. \label{fig:s1.pdf}}
\end{figure}
Algebraically we can write,
\begin{align}
\alpha_0=\langle 1 | e^{i \tilde\varphi (L/2)} | 0 \rangle \langle 0 | e^{-i \tilde\varphi (0)} | 1 \rangle e^{-i\pi/3}.
\end{align}
Second term in Eq.\eqref{pert} can be represented with four different diagrams corresponding to parameters $\alpha_1,\beta,\gamma,\lambda$. For $\alpha_1$ we have the following Figure,
\begin{figure}[H]
\vspace*{0.1in}
\centering
\includegraphics[width=0.6\columnwidth,keepaspectratio]{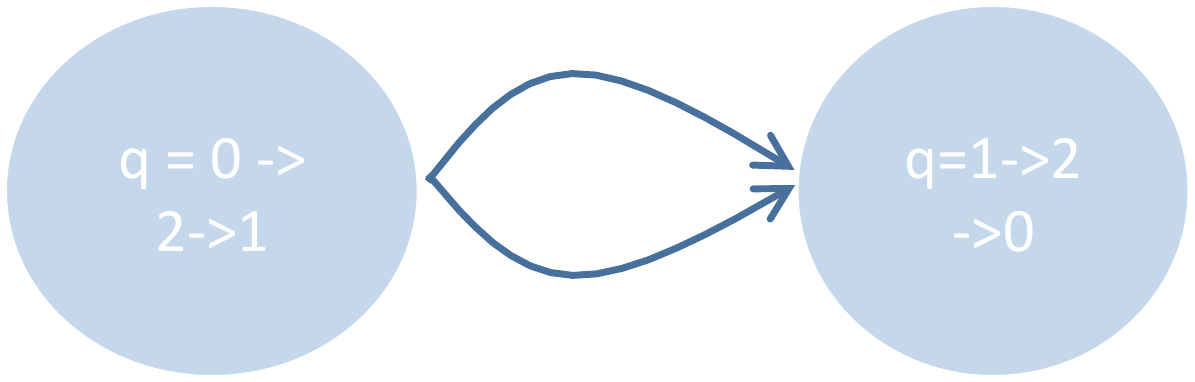} \caption{$\sigma^+_{i} \sigma^-_{i+1}$ term as a second order process in perturbation theory. $q$ is the fractional charge modulo three. \label{fig:s2.pdf}}
\end{figure}
note that this term also breaks conservation of fractional charge and therefore is only allowed at nonzero $\Delta$. In algebraic form we have,
\begin{align}
\alpha_1=e^{2i\pi/3} \sum_{j,j'}&\frac{\langle 1 | e^{-i \tilde\varphi (L/2)} | 2^{(j)} \rangle \langle  2^{(j)} | e^{-i \tilde\varphi (L/2)} | 0 \rangle}{E(2^{(j)})+E(2^{(j')})} \\ \newline \nonumber \times& \langle 0 | e^{i \tilde\varphi (0)} | 2^{(j')} \rangle \langle  2^{(j')} | e^{i \tilde\varphi (0)} | 1 \rangle
\end{align}
where $| 2^{(j)} \rangle$ corresponds to $j$'th state (arbitrary ordering) with $q=2$, and the energy $E$ is the bare $\Delta=0$ energy of the state with respect to the ground state energy. For $\beta$ we have the diagram,
\begin{figure}[H]
\centering
\includegraphics[width=\columnwidth,keepaspectratio]{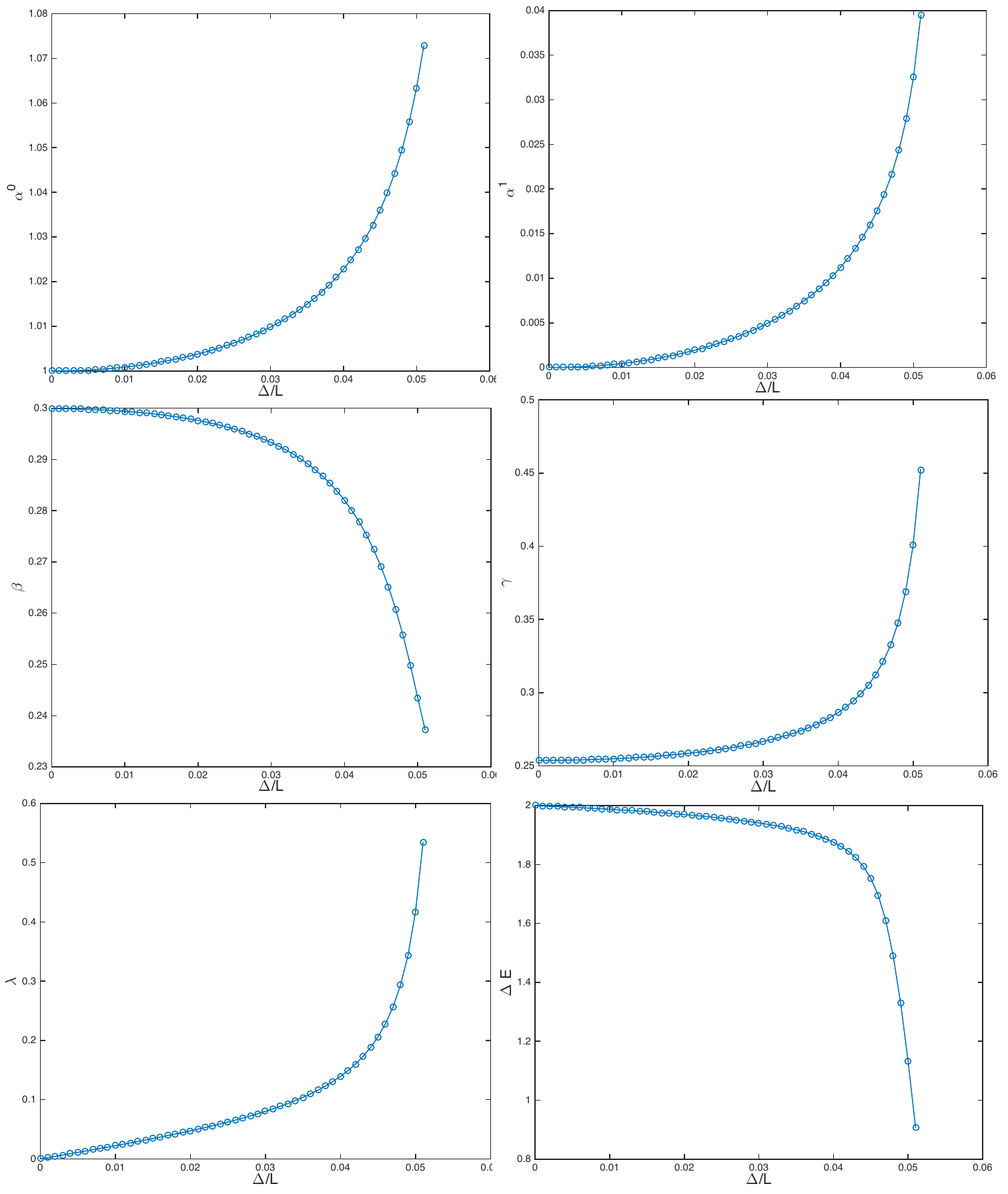} \caption{Parameters $\alpha_0,\alpha_1,\beta,\gamma,\lambda,\Delta E$, as a function of $\Delta$. Assuming ``pseudo point-like" superconductivity with $K{max}=4$ and $ m_\mu=\frac{u\pi\nu}{L}=1$. \label{fig:new11.pdf}}
\end{figure}
\begin{figure}[H]
\vspace*{0.1in}
\centering
\includegraphics[width=0.6\columnwidth,keepaspectratio]{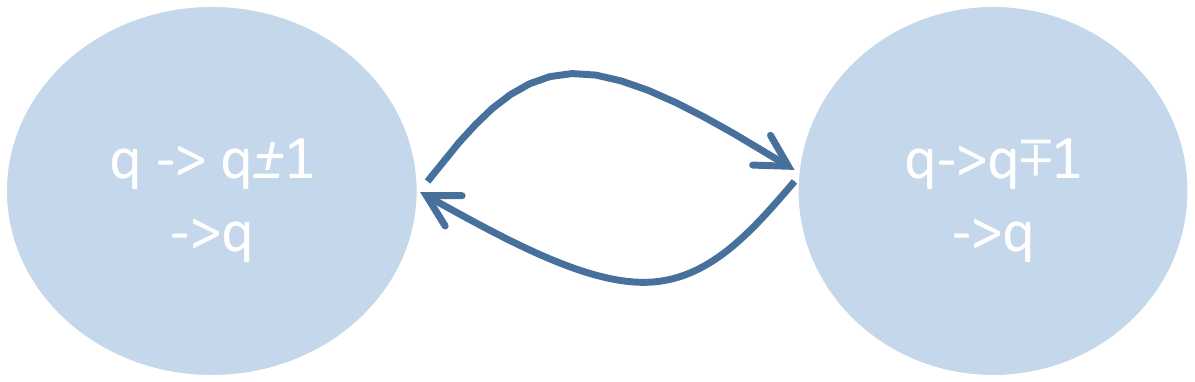} \caption{$\sigma^z_{i} \sigma^z_{i+1}$ term as a second order process in perturbation theory. $q$ is the fractional charge modulo three. \label{fig:s3.pdf}}
\end{figure}
Using $\mathbb{Z}_2$ symmetry we can write,
\begin{align}
&\beta=\sum_{j,j'}\frac{\langle 1 | e^{-i \tilde\varphi (L/2)} | 2^{(j)} \rangle \langle  2^{(j)} | e^{i \tilde\varphi (L/2)} | 1 \rangle}{E(2^{(j)})+E(0^{(j')})} \\ \newline \nonumber \times& \langle 1 | e^{i \tilde\varphi (0)} | 0^{(j')} \rangle \langle  0^{(j')} | e^{-i \tilde\varphi (0)} | 1 \rangle +  \langle 1 | e^{-i \tilde\varphi (0)} | 2^{(j')} \rangle \langle  2^{(j')} | e^{i \tilde\varphi (0)} | 1 \rangle \\ \newline \nonumber \times& \frac{\langle 1 | e^{i \tilde\varphi (L/2)} | 0^{(j)} \rangle \langle  0^{(j)} | e^{-i \tilde\varphi (L/2)} | 1 \rangle}{E(2^{(j')})+E(0^{(j)})}
\end{align}
where $| 2^{(j)} \rangle,| 0^{(j)} \rangle$ correspond to all excited states with $q=0,2$. Diagram corresponding to $\gamma$ term is,
\begin{figure}[H]
\vspace*{0.1in}
\centering
\includegraphics[width=\columnwidth,keepaspectratio]{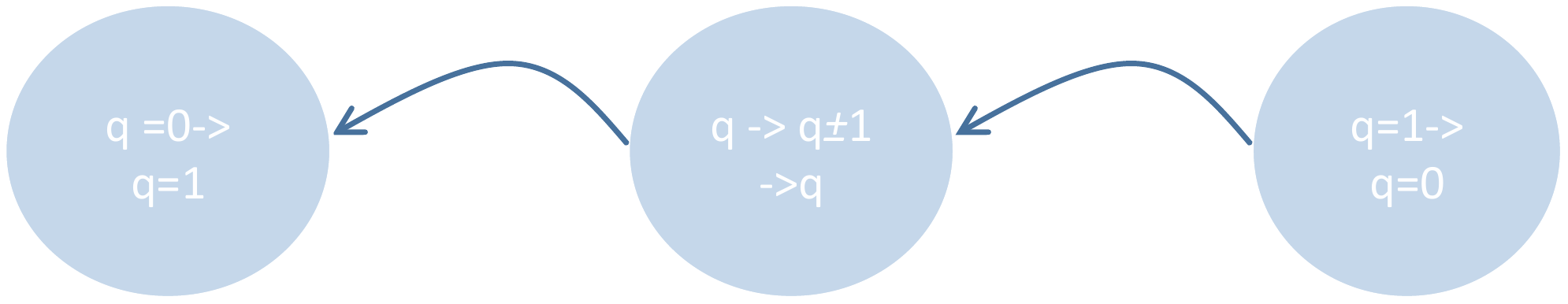} \caption{$\sigma^+_{i-1} \sigma^-_{i+1}$ term as a second order process in perturbation theory. $q$ is the fractional charge modulo three. \label{fig:s4.pdf}}
\end{figure}
We have,
\begin{align}
\gamma=&2\alpha_0 e^{-i\pi/3}\sum_j \Big( \frac{\langle 0 | e^{-i \tilde\varphi (0)}  | 1^{(j)} \rangle \langle 1^{(j)} | e^{i \tilde\varphi (L/2)}  |0\rangle}{E(1^{(j)})}\\ \newline \nonumber  &- \frac{e^{-i\pi/3} \langle 0 | e^{i \tilde\varphi (L/2)}  | 2^{(j)} \rangle \langle 2^{(j)} | e^{-i \tilde\varphi (0)}  |0\rangle}{E(2^{(j)})}\Big).
\end{align}
For the $\lambda$ term, corresponding diagram is given in Fig.~\ref{fig:pic3.pdf}. Algebraic form of $\lambda$ is,
\begin{align}
\lambda&=\langle 0 | e^{-i \tilde\varphi (L/2)} | 1 \rangle \langle 0 | e^{-i \tilde\varphi (0)} | 1 \rangle \sum_j \Big( \frac{e^{-4i\pi/3}}{E(2^{(j)})}\\ \newline \nonumber&\times
\langle 0 | e^{i \tilde\varphi (L/2)} | 2^{(j)} \rangle \langle 2^{(j)} | e^{i \tilde\varphi (0)} | 1 \rangle +\frac{e^{-2i\pi/3}}{E(2^{(j)})} \\ \newline \nonumber &\times
\langle 0 | e^{i \tilde\varphi (0)} | 2^{(j)} \rangle \langle 2^{(j)} | e^{i \tilde\varphi (L/2)} | 1 \rangle  \Big).
\end{align}
In Fig~\ref{fig:new11.pdf} we provide a plot of parameters $\alpha_0,\alpha_1,\beta,\gamma,\lambda,\Delta E$ ($\Delta E$ is the energy gap) as function of $\Delta$. We work with ``pseudo point-like" superconductivity with $K{max}=4$ and $ m_\mu=\frac{u\pi\nu}{L}=1$.

\section{Three fold degenracy in large $\Delta$ regime for point-like superconductivity}
\subsection{I. Diagonalizing the Hamiltonian in the limit $\Delta\rightarrow\infty$}

We study the Hamiltonian  $H=H_{edge}+H_{sc}$ in the limit of strong and point-like superconductivity. Note that the Hamiltonian commutes with the  operator $T = e^{-i\pi\nu\hat{n}}$, $[H,T] = 0$. Eigenvalue of $T$ can be shifted by a unitary transformation $e^{i\hat{\varphi}_0}$ as can be seen from the commutation $T e^{i{q}\hat{\varphi}_0}=e^{i{q}\hat{\varphi}_0}T e^{-i{q}\pi\nu}$,
i.e. if we define $H_{{q}=0}$ as the Hamiltonian $H$ in the charge sector ${q}=0$,  other charge eigenvalues
${q}$ can be generated from the transformed Hamiltonian,
\begin{align}\label{am1}
&H_q( m_\mu)=e^{-i{q}\hat{\varphi}_0}H_{{q}=0}( m_\mu +q)e^{i{q}\hat{\varphi}_0}.
\end{align}

We emphasize that $H_{{q}=0}$ is defined in the Hilbert space where the wavefunctions are eigenstates of $T$ with eigenvalue $1$ which in turn implies that the wavefunctions
are periodic under the translation $\varphi_0\rightarrow \varphi_0+\pi \nu$ (same as saying allowed charges are multiples of three ).

In the range of periodicity of the wave-functions, the potential $H_{sc}$ has a single minimum. In the limit of large $\Delta$, $H_{sc}$ we can expand around this minimum to obtain the harmonic approximation of $H_{sc}$,
\begin{align}
H_{sc}\sim \Delta (1-\frac{9}{2}\varphi^2(0)) .
\label{perturbation}
\end{align}

Note that the continuous nature of  the harmonic approximation means that wave-functions would in general violate the periodicity condition as we change $\varphi_0\rightarrow \varphi_0+\pi\nu$. However, this boundary condition is expected to be irrelevant, for calculating ground state energies, in large delta regime where $\varphi_0$ becomes strongly localized around zero. In the next section, we'll use an instanton approximation to see how enforcing periodic boundary conditions modifies our results.

Within the harmonic approximation $H_{{q=0}}$ can be turned into a set of decoupled harmonic oscillators with trivial spectrum. A particularly nice feature of this transformation is that the spectrum of $H_{{q}}$ will not depend of ${q}$ (spectrum of $H_{q=0}$ will not depend on $ m_\mu$). Degeneracy of the three charge sectors in the large $\Delta$ regime follows from this. We'll now show this by explicitly diagonalizing the Hamiltonian.

To diagonalize the Hamiltonian, it's useful to define the following generalized ``position" and ``momentum" operators,
\begin{align}
\hat{\beta}_k \equiv i\sqrt{\frac{L}{2u\pi k}}\left(\hat{a}_k -\hat{a}_k^\dagger\right) \quad ; \quad
\hat{\alpha}_k \equiv \sqrt{\frac{u\pi k}{2L}}\left(\hat{a}_k +\hat{a}_k^\dagger\right)
\end{align}
These operators satisfy,
\begin{align}
[\hat{\beta}_k,\hat{\alpha}_{k'}] = i\delta_{kk'}  \quad ; \quad
[\hat{\beta}_k,\hat{\beta}_{k'}] = [\hat{\alpha}_k,\hat{\alpha}_{k'}] =0.
\end{align}
In these variables and within harmonic approximation $H_{q=0}$ becomes (ignoring constants),
\begin{align}
&H_{q=0,h}= \hat{\mathrm{X}}^\mathrm{T}  H_{\mathrm{X}}  \hat{\mathrm{X}} + \hat{\mathrm{P}}^\mathrm{T}\hat{\mathrm{P}}  \qquad\qquad \mathrm{with,}\nonumber \\
\hat{\mathrm{X}} \equiv &\left(\sqrt{\frac{L}{u \pi \nu}}\hat{\varphi}_0,\hat{\beta} \right)  ;
\hat{\mathrm{P}} \equiv \left(\sqrt{\frac{u\pi \nu}{L}}(\hat{n}- m_\mu),\hat{\alpha}\right).
\label{matrixH}
\end{align}

Where(setting $ m_\mu =1$),
\begin{align}
H_{\mathrm{X}}=\frac{9u\Delta\pi\nu}{2L} \begin{pmatrix}
1 & 1/\sqrt{2} \\
1/\sqrt{2} & 1+\frac{2 u\pi k^2}{9L\Delta \nu} \delta_{k,k'}
\end{pmatrix}
\end{align}

Note that since the commutator $[\hat{\varphi_0},\hat{n}- m_\mu]=i$ is independent of $ m_\mu$, the spectrum of $H_{q=0,h}$ does not depend on $ m_\mu$. Which in turn implies that the spectrum of $H_q$ does not depend on $q$

\begin{figure}[t]
 \begin{center}
\includegraphics[width=\columnwidth]{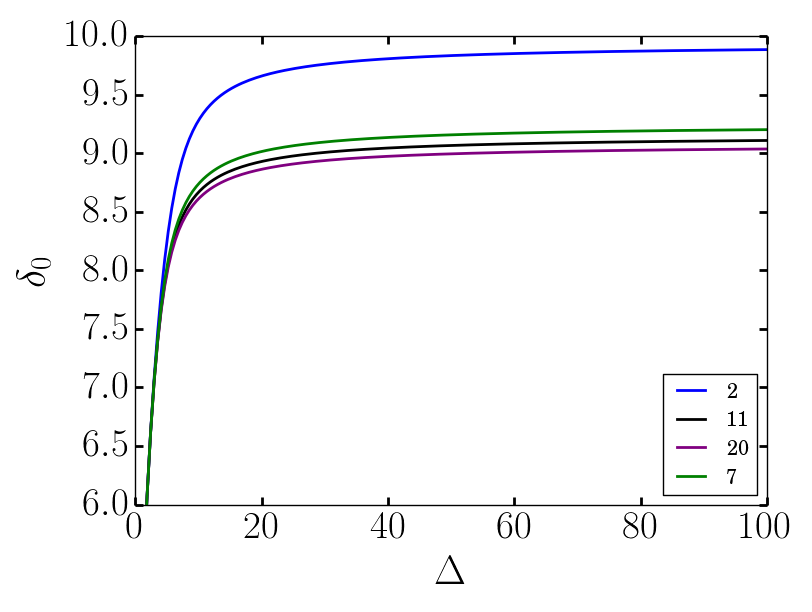}
\end{center}
\caption[]{
Plot of the energy gap $\delta_0$ as a function of $\Delta$ for various total number of modes $K_{max}$ shown in caption. Here numerical values of $ m_\mu$, $u$ and $L$ are set to one.
}
\label{E_GS}
\end{figure}

In its diagonal form this Hamiltonian describes a set of decoupled harmonic oscillators with different frequencies. In Fig.~\ref{E_GS} we have plotted the energy gap $\delta_0$ as a function of $\Delta$.

\begin{figure}[t]
 \begin{center}
\includegraphics[width=\columnwidth]{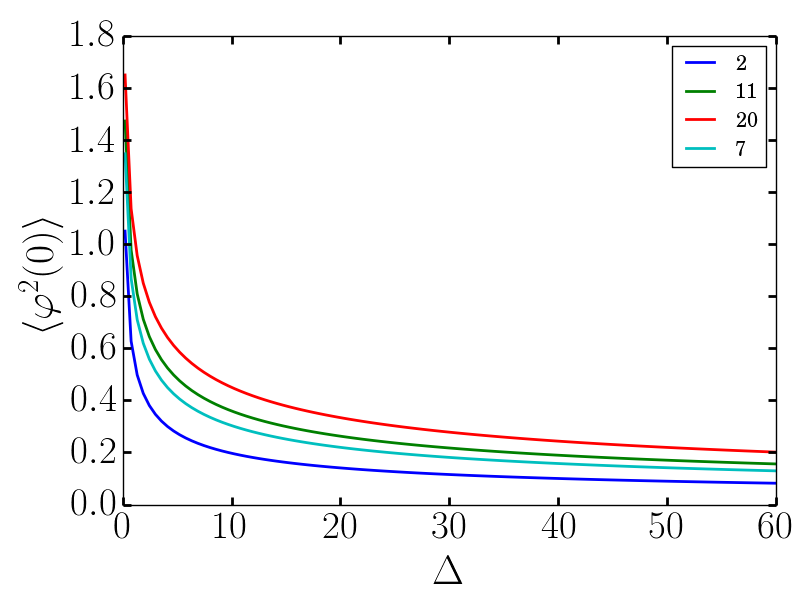}
\end{center}
\caption[]{$\langle\varphi^2(0)\rangle$ as a function of $\Delta$ for various total number of modes $K_{max}$ shown in caption. We find $\varphi (0)$ is localized for large values of $\Delta$. Therefore, for large enough value of $\Delta$, harmonic approximation is justified. Here numerical values of $ m_\mu$, $u$ and $L$ are set to one. }
\label{localized}
\end{figure}

This completes our discussion of the spectrum within harmonic approximation.

As a measure of validity for our approximation, we calculate the expectation value of $\langle\varphi^2(0)\rangle_0$ . As shown in Fig.~\ref{localized} we find that the fluctuations are monotonically decreasing function of the coupling strength $\Delta$. In other words, one can always choose the value of $\Delta$ to be large enough to get phase fluctuations in $\varphi(0)$  to be much smaller than $2\pi$ which justifies the use of harmonic approximation.

\subsection{II. Finite $\Delta$ corrections to ground state energies}
\begin{figure}[t]
	\begin{center}
		\includegraphics[width=\columnwidth]{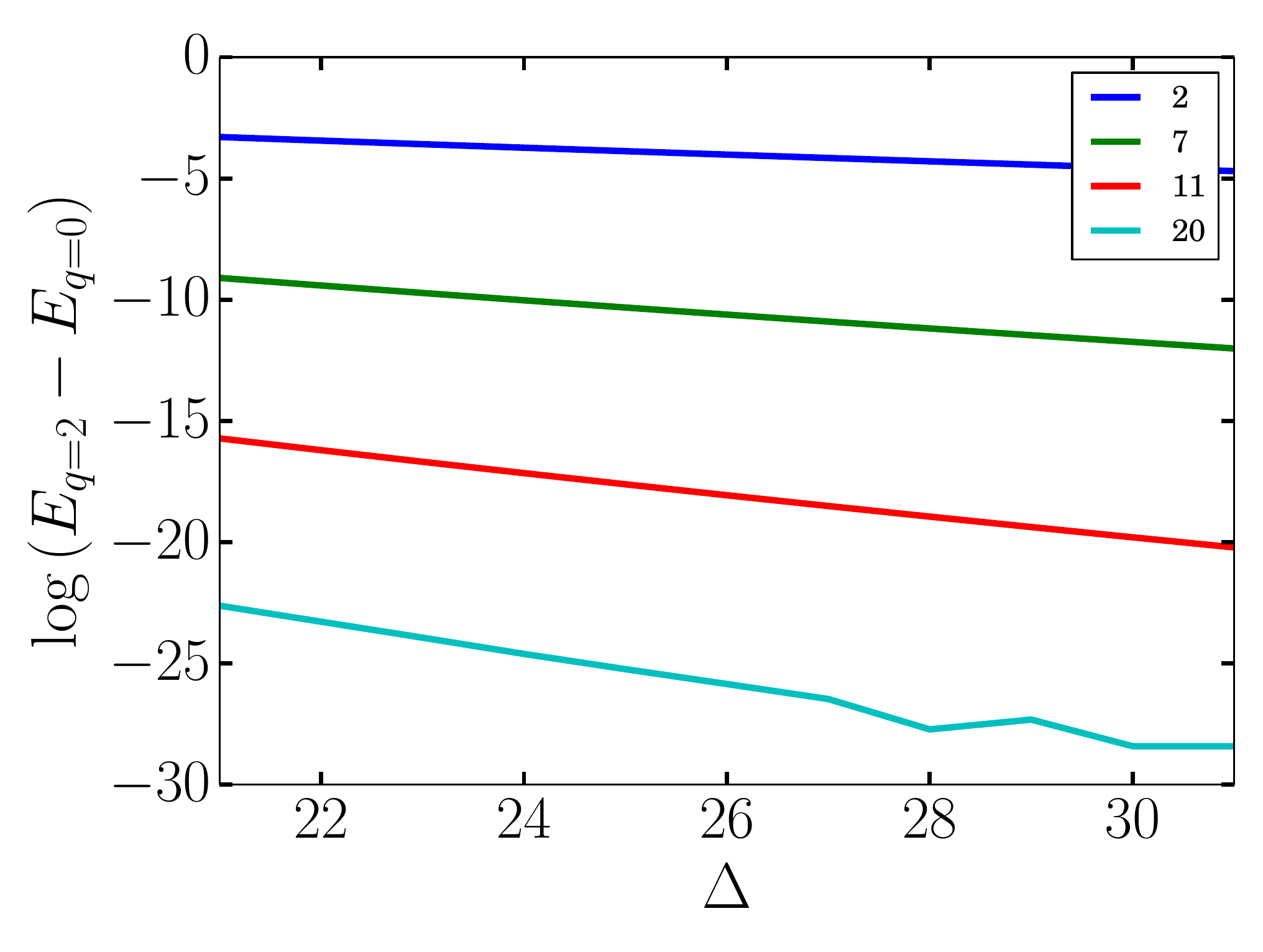}
	\end{center}
	\caption[]{
		Logarithm of ground-state energy splitting between charge sectors ${q} = 2$ and ${q}=0,1$ (they're degenerate) as a function of coupling strength $\Delta$,  for different total number of modes $K_{max}$ shown in the inset. Notice that the energy splitting goes to zero at large $\Delta$. Here numerical values of $ m_\mu$, $u$ and $L$ are set to one.}
	\label{Instanton}
\end{figure}

As mentioned earlier wave functions found within harmonic approximation violate the periodicity condition as we change $\varphi_0\rightarrow \varphi_0+\pi\nu$. In this section we use an instanton approximation to see how this periodicity changes our previous results.

We'll enforce periodicity by externally projecting the states into the physical Hilbert space,
\begin{align}
&|{q=0}\rangle=\mathcal{P}|{q=0}\rangle_h,
\end{align}
where $|{q=0}\rangle_h$ is the $q=0$ eigenstate within the harmonic approximation and $\mathcal{P}$ is an
operator that projects into the sector that obeys the  periodicity condition as $\varphi_0\rightarrow \varphi_0+\pi\nu$. We then calculate the ${q}$ and $\Delta$ dependence of ground-state energy using,
\begin{align}
E_{{q}} &= \langle {q}|H|{q}\rangle/\langle {q}|{q}\rangle \nonumber \\
&= \frac{\sum_{m\in\mathbb{Z}} e^{-i\pi\nu{q}m}N(m)}{\sum_{m\in\mathbb{Z}}e^{-i\pi\nu{q}m} D(m)},
\end{align}
where $N(m)= N_1(m) + N_2(m) + N_3(m) + N_4(m)$ with the definitions,
\begin{align}
N_1(m) &=  \left(\sum_p(\lambda_+^{0n})^2-\left(\pi\nu m\sum_p(\lambda_+^{0n})^2\right)^2\right)\prod_ne^{-(\frac{1}{2}h_{nm})^2} \nonumber \\
N_2(m) &= -\prod_ne^{-(\frac{1}{2}h_{nm})^2}\left(\sum_k \frac{2\pi uk}{L}\left( \sum_n(\lambda_-^{kn})^2 \right.\right.\nonumber \\
&+\left.\left. (\pi\nu m)^2\left(\sum_p\lambda_+^{kp}\lambda_+^{0p}\right)\left(\sum_q\lambda_-^{kq}\lambda_+^{0q}\right)\right)\right) \nonumber \\
N_3(m) &= \left(\sum_k \frac{u\pi k}{L}\right)\prod_ne^{-(\frac{1}{2}h_{nm})^2} \nonumber \\
N_4(m) &= \frac{9\Delta}{2} \left(\prod_n e^{-\frac{1}{2}\left(h_{nm}^2-f_{n}^2\right)-h_{nm}f_{n}}\right.\nonumber \\
&\left.+ \prod_n e^{-\frac{1}{2}\left(h_{nm}^2-f_{n}^2\right)+h_{nm}f_{n}}\right) \nonumber \\
D(m) &= \prod_ne^{-(\frac{1}{2}h_{nm})^2},
\end{align}
with,
\begin{align}
f_n &= \frac{2}{\nu}\left(\sum_k\sqrt{\frac{\nu}{k}}(\lambda_-^{kn}+\lambda_+^{kn})+\lambda_-^{0n}\right) \nonumber \\
h_{nm} &= \pi\nu m\lambda_{+}^{0n} \nonumber \\
\lambda_+^{0n} &= \frac{1}{2}\sqrt{\frac{L}{u\pi\nu}}U^{\dagger}_{0n}\sqrt{\omega_n} \nonumber \\
\lambda_-^{0n} &= i\sqrt{\frac{u\pi\nu}{L}}U_{0n}\frac{1}{\sqrt{\omega_n}} \nonumber \\
\lambda^{kn}_+ &= \frac{i}{2}\sqrt{\frac{L}{2\pi uk}} U^{\dagger}_{kn}\sqrt{\omega_n} + \frac{i}{2}\sqrt{\frac{2\pi uk}{L}} U_{kn}\frac{1}{\sqrt{\omega_n}} \nonumber \\
\lambda^{kn}_- &= -\frac{i}{2}\sqrt{\frac{L}{2\pi uk}} U^{\dagger}_{kn}\sqrt{\omega_n} + \frac{i}{2}\sqrt{\frac{2\pi uk}{L}} U_{kn}\frac{1}{\sqrt{\omega_n}}
\end{align}
where, $U_{mn}$ and $\omega_n$ are the matrix element of unitary matrix $U$, diagonalizing $H_X$ and the $n^{th}$ eigenvalue of $H_h$, respectively (see Eq.~\eqref{matrixH}).

Sample results of this calculation are shown in Fig.~\ref{Instanton}. This results can be used as an estimate for the value of the parameter $h$ in the clock model Hamiltonian \eqref{sm4}.

\subsection{III. Quasiparticle matrix elements}
\begin{figure}
	\begin{center}
		\includegraphics[width=\columnwidth]{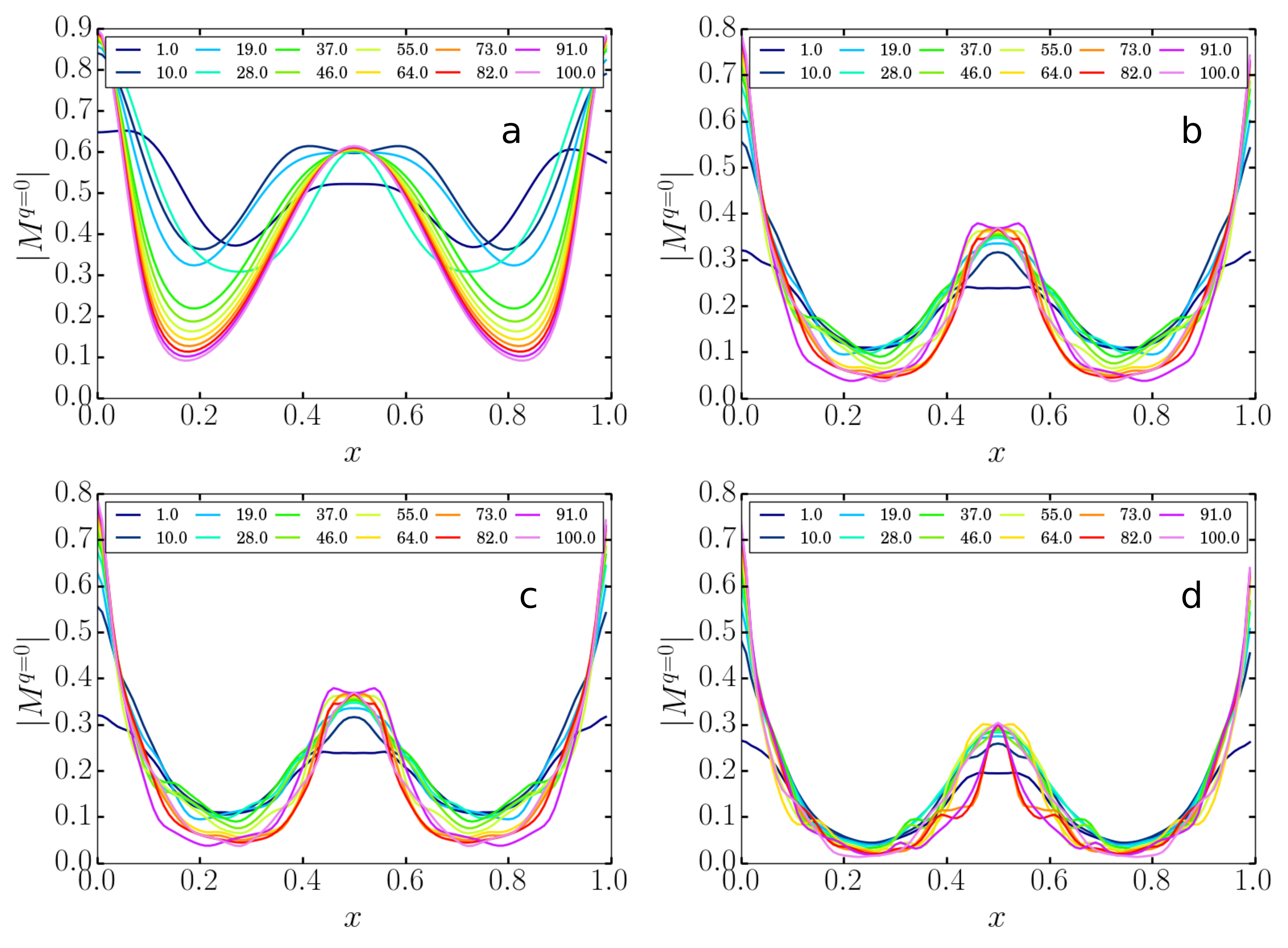}
	\end{center}
	\caption[]{
		Plot of absolute value of quasi-particle matrix element, $|M^{q=0}|$ as a function x (in the units of length $L$) for various coupling strength parameters, $\frac{4}{\nu^2}\Delta$, given in the legend. The total number of modes $K_{max}$ are varied in (a-d) as 2,7,11 and 20, respectively. Here numerical values of $ m_\mu$, $u$ and $L$ are set to one.}
	\label{QPmatrixElementQ0}
\end{figure}

\begin{figure}
	\begin{center}
		\includegraphics[width=\columnwidth]{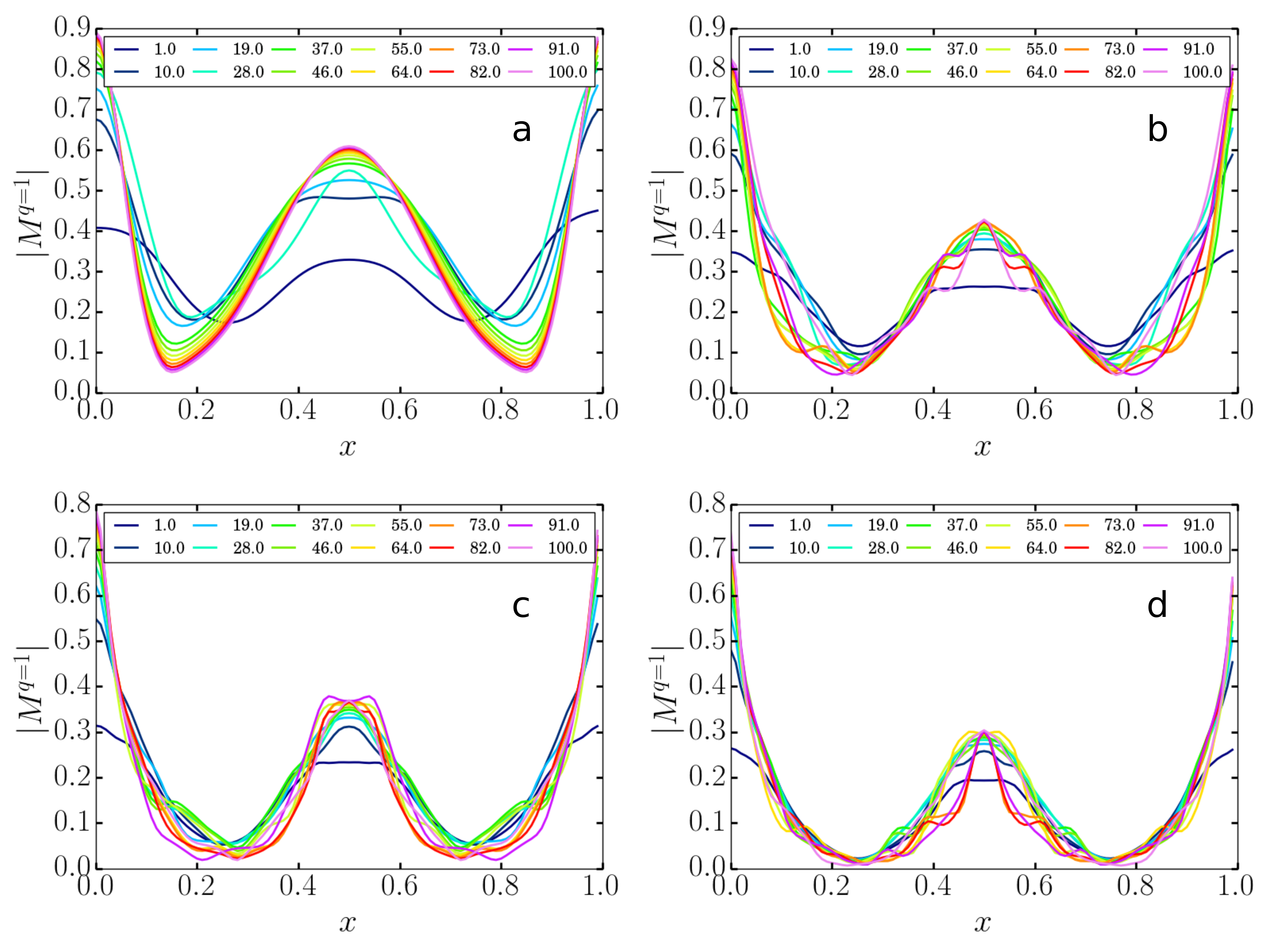}
	\end{center}
	\caption[]{
		Plot of absolute value of quasi-particle matrix element, $|M^{q=1}|=|M^{q=2}|$ as a function x (in the units of length $L$) for various coupling strength parameters, $\frac{4}{\nu^2}\Delta$, given in the legend. The total number of modes $K_{max}$ are varied in (a-d) as 2,7,11 and 20, respectively. Here numerical values of $ m_\mu$, $u$ and $L$ are set to one.}
	\label{QPmatrixElementQ1}
\end{figure}

Let us consider quasiparticle tunneling elements between the ground states ,
\begin{align}
&M^{{q}}(x)=\langle {q}+1|e^{i\varphi(x)}|{q}\rangle.
\end{align}
In terms of states in the harmonic approximation,
\begin{align}
&M^{{q}}(x)=\langle {q}+1|_h\mathcal{P}^{\dagger}e^{-i({q}+1)\hat{\varphi}_0}e^{i\varphi(x)}e^{i{q}\hat{\varphi}_0}\mathcal{P}|{q}\rangle_h.
\end{align}
Setting $ m_\mu=1$ and substituting $\mathcal{P} = \mathcal{N}\sum_me^{i\pi\nu m\hat{n}}$ ($\mathcal{N}$ being the normalization constant) in the harmonic approximation for $|{q}\rangle=e^{-i{q}\hat{\varphi}_0}|0\rangle$, we get,
\begin{align}
M^{{q}} &=\mathcal{N}^2\sum_{p,m\in\mathbb{Z}} e^{i\pi\nu(p-m)({q}+1)}\langle 0|e^{i\pi\nu\hat{n}(m-p)}e^{i\varphi (x)}|0\rangle \\ \nonumber
&\equiv \sum_{m\in\mathbb{Z}} M_m^{{q}}
\end{align}
with,
\begin{align}
&\quad M_m^{{q}} = e^{i\pi\nu m({q}+1)}\langle 0|e^{-i\pi\nu m\hat{n}}e^{i\varphi(x)}|0\rangle = e^{i\pi\nu m({q}+1/2)} \nonumber \\
&\times\langle 0|e^{i(\hat{\varphi}_0+\pi\nu(2x/L-m)\hat{n}+i\sum_k\sqrt{\frac{\nu}{k}}(\hat{a}_ke^{i\frac{2\pi kx}{L}}-\hat{a}_k^{\dagger}e^{-i\frac{2\pi kx}{L}})}|0\rangle
\end{align}
This explicit form makes clear the following identity,
\begin{align}
M_{m+2}^{{q}}(x+L) &= e^{i\pi\nu m(2{q}+1)}M_m^{{q}}(x) \nonumber \\
\implies M^{{q}}(x+L) &= e^{i\pi\nu m(2{q}+1)}M^{{q}}(x)
\end{align}
Using the expressions,
\begin{align}
\hat{a}_k^{\dagger} &= \sum_n(\lambda^{kn}_+ A_n^{\dagger} + \lambda^{kn}_- A_n) \nonumber \\
\hat{a}_k &= -\sum_n(\lambda^{kn}_- A_n^{\dagger} + \lambda^{kn}_+ A_n)\nonumber \\
\hat{\tilde{n}}_{{q}=0} &= \sqrt{\frac{L}{\pi\nu u}}\frac{-i}{2}\sum_n U^{\dagger}_{0n}\sqrt{\omega_n}(A_n - A_n^{\dagger}) \nonumber \\
\hat{\varphi}_0 &= \sqrt{\frac{\pi\nu u}{L}}\sum_n U_{0n}\frac{1}{\sqrt{\omega_n}}(A_n + A_n^{\dagger}),
\end{align}
where $A_n$ is the $n^{th}$ component of the operator valued vector $A \equiv \frac{1}{\sqrt{2}}\left(\hat{\mathrm{X}} + i\hat{\mathrm{P}}\right)$, $M_m^{{q}}$ can be expressed as,
\begin{align}
M_m^{{q}} = e^{i\pi\nu m{q}}\prod_n e^{-\frac{1}{2}|\eta_n^m(x)|^2}\langle 0|e^{-\eta^{m*}_nA_n^{\dagger}}e^{\eta_n^mA_n}|0\rangle
\end{align}
with,
\begin{align}
&\eta^m_n(x) = \left(2\pi\nu\frac{x}{L}-\pi\nu m\right)\frac{1}{2}\sqrt{\frac{L}{\pi\nu u}}U_{0n}^{\dagger}\sqrt{\omega_n} \nonumber \\
&+i\sqrt{\frac{\pi\nu u}{L}}U_{0n}\sqrt{\frac{1}{\omega_n}}+\sum_k\sqrt{\frac{\nu}{k}}(\lambda_+^{kn}e^{-i2\pi kx/L}+\lambda_-^{kn}e^{i2\pi kx/L}) .
\end{align}
We use this expression to calculate $|M^{q}|$ as a function of number of modes and coupling strength, $\Delta$. Since $|M^{q=1}|=|M^{q=2}|$, we plot the results for $|M^{q=0}|$ and $|M^{q=1}|$. The results are shown in Figs.~\ref{QPmatrixElementQ0} and ~\ref{QPmatrixElementQ1}.

These results can be used to calculate the parameter $A(x)$ in Eq.\eqref{sm2}.

\end{document}